\documentclass{emulateapj}

\def\hi{\ifmmode {\rm H}\,{\sc i}~ \else H\,{\sc i}~\fi}

\def\chandra {\emph{Chandra}}

\slugcomment{Accepted to ApJ Letters}

\shorttitle{A galaxy filament associated with WHIM absorption}
\shortauthors{Williams et al.}

\begin{document}

\title{Discovery of a large-scale galaxy filament near a candidate 
intergalactic X-ray absorption system}

\author{Rik J. Williams, John S. Mulchaey, Juna A. Kollmeier, and Thomas J. Cox}
\affil{Carnegie Observatories, 813 Santa Barbara St., Pasadena, 
CA 91101, USA}

\email{williams@obs.carnegiescience.edu}

\begin{abstract}
We present an analysis of the large-scale galaxy distribution around two
possible warm-hot intergalactic medium (WHIM) absorption 
systems reported along the Markarian 421 sightline.  Using the Sloan 
Digital Sky Survey, we find a prominent galaxy
filament at the redshift of the $z=0.027$ X-ray absorption line system.
The filament exhibits a width of 3.2 Mpc and length of at least
20 Mpc, comparable to the size of WHIM filaments seen in cosmological 
simulations.  No individual galaxies fall within 350 projected kpc so it is 
unlikely that the absorption
is associated with gas in a galaxy halo or outflow.  
Another, lower-significance X-ray absorption system was reported in the 
same \chandra\ spectrum at $z=0.011$, but the large-scale 
structure in its vicinity is far weaker and may be a spurious alignment.  
By searching for similar galaxy structures in 140 random smoothed SDSS 
fields, we estimate a $\sim 5-10$\% probability of the $z=0.027$ 
absorber-filament alignment occurring by chance.
If these two systems are  indeed physically associated, this 
would represent the first known coincidence between large-scale galaxy 
structure and a blind X-ray WHIM detection.  
\end{abstract}


\section{Introduction}
Despite theoretical expectations that the warm-hot ($T>10^5$\,K) component
of the $z<1$ intergalactic medium contains $25-50$\% of the baryons
in the local universe \citep{dave10}, this material has thus far frustrated 
attempts
at unambiguous detection and characterization.  This is primarily due to
its high temperatures ($10^5-10^7$ K) and low densities
($10^{-6} - 10^{-4}$ cm$^{-3}$), rendering it all but undetectable through
well-studied UV/optical absorption lines like Lyman-$\alpha$, Mg\,{\sc ii},
and oftentimes even O\,{\sc vi}.  Instead, the primary 
signature of 
this warm-hot intergalactic medium (WHIM) is expected to be highly-ionized 
metal X-ray lines \citep[both in absorption against bright background quasars 
and diffuse emission;][]{cen99}, most notably O\,{\sc vii}, O\,{\sc viii}, and 
Ne\,{\sc ix}.  

Given the weakness of the signal, detections of the WHIM have thus far been 
highly uncertain and/or ambiguous.  A single $z=0.055$ O\,{\sc viii} line 
toward PKS 2155-304 (possibly associated with a small galaxy group) has 
been repeatedly confirmed 
\citep{fang02,fang07,williams07}, but without other detected lines at
the same redshift this result remains uncertain.  
\citet[][hereafter N05]{nicastro05a,nicastro05b}
reported ``blind'' detections of two WHIM lines in an extremely
high signal-to-noise spectrum of the blazar Mrk 421 in outburst, with
one possible system at $z=0.011$ and another detected in multiple ionic species
at $z=0.027$.  However, this result proved controversial given their
non-detection with the XMM-Newton Reflection
Grating Spectrometer; statistical re-analyses of the absorption systems'
significance have also called their veracity into question 
\citep{kaastra06,rasmussen07}.  Although the 
reported XMM upper limits are 
formally consistent with the best-fit Chandra measurements, and 
certain instrumental characteristics of the XMM RGS can seriously hinder 
weak absorption line studies \citep{williams06}, this controversy nonetheless
highlights the extreme difficulty of such blind WHIM searches.

Other studies have found stronger detections of possible WHIM lines, but
the interpretation remains ambiguous.  \citet{williams05} performed
detailed ionization and curve-of-growth analyses of the strong $z=0$
X-ray absorption in the Mrk 421 spectrum; while these X-ray lines 
are inconsistent with \emph{known} Galactic gaseous components, an origin in
the large-scale WHIM surrounding the Galaxy and Local Group could not 
be conclusively inferred.  Emission signals detected indirectly via the 
X-ray background autocorrelation function \citep{galeazzi09} and 
directly in overdense regions \citep{zappacosta05,mannucci07,werner08}
have also been found, but the former method relies heavily on proper 
point-source
subtraction and the latter may be probing extended, bound intracluster 
gas rather than the ``representative'' diffuse WHIM. 
 
Perhaps the most convincing evidence thus far has been the detection of
X-ray absorption lines associated with large-scale, nonvirialized
galaxy structures like the Sculptor Wall \citep{buote09,fang10} 
and the Pisces-Cetus Supercluster \citep{zappacosta10}.  Such filamentary
structures are expected to be associated with large reservoirs of diffuse
WHIM gas \citep{dave01}, and thus X-ray absorption searches focusing on such
systems have a higher probability of success than their blind counterparts.
However, this method also has its drawbacks: notably, gas 
associated with these moderately high-density regions may still not be 
representative of typical WHIM properties, and individual galaxies 
with small ($\la 50$\,kpc) impact parameters can contaminate the WHIM 
absorption (J.~Mulchaey et al., in preparation), thereby complicating the
interpretation of a detection.

Nonetheless, the spatial coincidence of even weak X-ray absorption lines
with other structures (whether galaxies, cluster outskirts, or large-scale
nonvirialized filaments) is exciting, since 
little is known about the warm-hot gas in these environments.
Here we revisit the \chandra-detected absorption line systems reported 
by N05, searching for
galaxy structures in public data that have been released since the
detections.  In \S\ref{sec_data}  we describe the data and selection procedure, 
in \S\ref{sec_filaments} we discuss the galaxies and structures found near 
the absorbers, interpretations of the results can be found in \S\ref{sec_discussion}, 
and we briefly summarize the conclusions in \S\ref{sec_summary}.  
Cosmological parameters $H_0 = 70$\,km s$^{-1}$ Mpc$^{-1}$, $\Omega_0=0.3$, 
and $\Omega_\Lambda=0.7$ are assumed throughout.

\section{Data and galaxy samples} \label{sec_data}
Markarian (Mrk) 421 is an X-ray bright $z=0.03$ blazar located
at $\alpha=11^{\rm h}04^{\rm m}27\fs 3, \delta=+38^{\rm d}12^{\rm m}32^{\rm s}$.
By combining two deep \chandra\ Low-Energy Transmission Grating (LETG) 
spectra of this object taken during extreme X-ray outbursts, N05 reported 
the detection
of two significant absorption line systems: one at $z=0.011\pm 0.001$ 
(detected in two absorption lines with $\geq 2\sigma$ significance)
and another at $z= 0.027\pm 0.001$ (detected in four lines, three of 
which are stronger than $2\sigma$).  
Hereafter we refer to these absorbers as ``A1'' and ``A2'' respectively.
The quoted redshift errors are primarily due to wavelength-dependent
uncertainties in the \chandra-LETG dispersion relation.  Taking
all detected lines together, N05 report a ``conservative'' significance
(i.e.~taking into account possible spurious line detections over a wide
wavelength range in their multi-component fits) of $3.5\sigma$ for A1 and 
$4.9\sigma$ for A2.

Since the publication of these putative WHIM detections, the Sloan
Digital Sky Survey \citep[SDSS;][]{york00} has publicly released a 
comprehensive catalog of galaxy redshifts spanning the northern Galactic 
cap (including the Mrk 421 region).  This survey is
at least 90\% complete to $r=17.77$ \citep{strauss02}, corresponding to
about $0.04 L_\star$ at $z=0.03$ \citep[where $L_\star$ is the
characteristic luminosity in the Schechter function;][]{blanton03}.   
Here the incompleteness is primarily due to the 55\arcsec\ minimum 
fiber spacing in the spectroscopic survey; in regions where the survey
tiles overlap the completeness is expected to be higher,
so 90\% should be considered a conservative lower limit.  In the SDSS
photometric catalog  only one $r<17.77$ galaxy lies within 55\arcsec\ of 
Mrk 421, a known companion to the quasar host at $z=0.03$.  Thus,
it is unlikely that any projected companions near Mrk 421
are missed due to the minimum fiber spacing.

In order to provide samples of galaxies and galaxy structures
both near and unrelated to the absorbers, we retrieved data on \emph{all} 
galaxies (unconstrained by sky position) from the SDSS Data
Release 7 \citep{abazajian09} ``BESTDR7'' database 
within $\pm 450$\,km\,s$^{-1}$ around the X-ray detections: 
$0.0095<z<0.0125$ and $0.0255<z<0.0285$.  These intervals are sufficiently 
broad to
encompass all of the detected lines' best-fit wavelengths in each absorption 
system, thus conservatively allowing for uncertainties in the absorbers'
redshifts (as well as possible systematic offsets between the galaxies
and filament gas) while effectively excluding foreground and background
interlopers.  Aside from the redshift constraints, 
we further required objects to be flagged as galaxies by the SDSS pipeline 
({\tt specClass} $==2$) and have high redshift confidence 
({\tt zConf} $> 0.95$).

\section{Galaxies and large-scale structures near the absorbers} \label{sec_filaments}
The line-of-sight galaxy redshift distribution in a narrow cylinder
(with a radius of 5\,Mpc) centered on Mrk 421 is shown in 
Figure~\ref{fig_zhist}; the mean absorber redshifts are marked with
arrows.  In both cases, the absorbers coincide with
significant spikes in the galaxy distribution: a rather broad one spanning 
approximately $z=0.009-0.012$ and a narrow one at $z=0.027$ (superposed
on a larger overdensity ranging from at least $z=0.025-0.033$).  
Interestingly, the space between the two absorbers is largely
devoid of galaxies.

\begin{figure}
\plotone{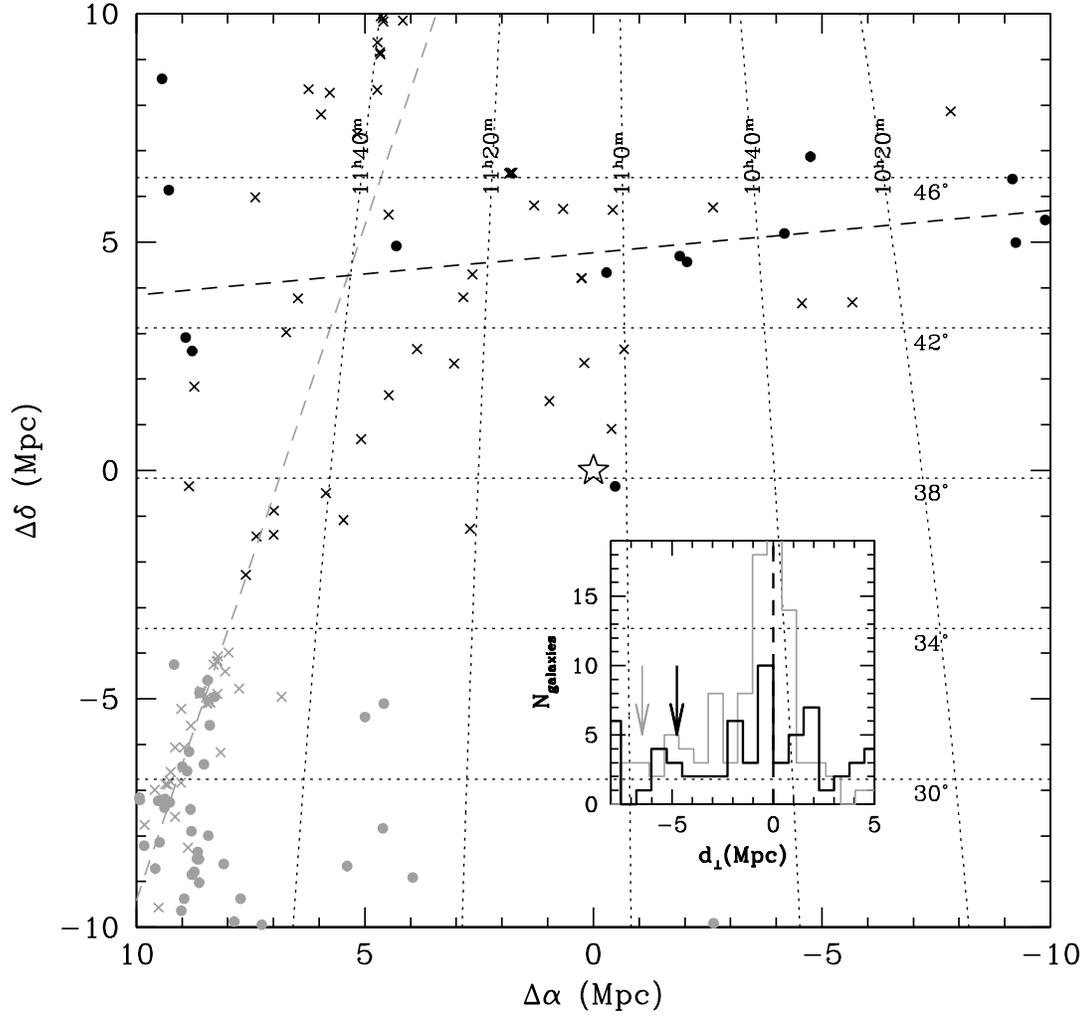}
\caption{Galaxies in a $20\times 20$ Mpc and $\Delta z = \pm 0.0015$  box 
around the $z=0.011$ X-ray absorption system (A1); north is up and east
is to the left.  Higher- and lower-velocity 
galaxies relative to the absorber are plotted as crosses and circles 
respectively, and the position of Mrk 421 is shown as a star.  The overlaid 
dotted grid denotes the SDSS J2000 coordinates.  The dashed black line traces 
the
approximate center of the nearest filament, calculated as a simple 
least-squares fit to the black points; grey points are not included in the 
fit (see \S\ref{sec_caveats}).  Another fit performed to the strong clump
of galaxies to the southeast is shown as a dashed grey line.
\emph{Inset:} The black (grey) histogram shows the distribution of 
galaxies' perpendicular
distance $d_\perp$ from the dashed black (grey) line;  the 
arrow at 4.7 (6.4)\,Mpc denotes
the absorber's distance.  Although a large-scale galaxy structure
is visible here, it is not particularly strong or close to the absorber.
\label{fig_fil011} }
\end{figure}

While the absorbers evidently lie within 5 projected Mpc of galaxy structures
in redshift space, the actual nature of these overdensities is not
obvious from the histogram.  
Figures~\ref{fig_fil011} and \ref{fig_fil027} therefore show the galaxy 
distributions within two narrow
redshift slices (depth $\Delta z=\pm 0.0015$) in boxes 20\,Mpc on a side. 
Objects blueshifted 
relative to the absorbers are plotted as solid circles, and redshifted
galaxies are shown as crosses.  Large-scale galaxy structures are
again apparent in both slices: near A1 there appears
to be a broad (albeit somewhat diffuse) filament or sheet of galaxies
to the north and east of the absorber, while A2 lies
on the edge of a strong, well-defined galaxy filament.  In both cases,
the absorbers appear to lie in ``boundary'' regions, near relatively
large voids or underdense regions.

To better quantify the structures near the absorbers, we reject by
eye galaxies that do not appear to be ``members'' of the structure
nearest each absorber (shown as grey points; this method is discussed 
further in \S\ref{sec_caveats}), 
and perform simple linear fits to the positions
of the remaining galaxies.  These fits are plotted as dashed lines in
Figures~\ref{fig_fil011} and \ref{fig_fil027}, and the number of galaxies
as a function of perpendicular distance from these best-fit lines are
shown in the inset plots.  The filament near A2 exhibits a strong,
nearly-Gaussian shape around this line with FWHM$=3.2$\,Mpc (though
non-Gaussian wings extending to $\pm 4$\,Mpc may also be present).
On the other hand, the $z=0.011$ structure
is not as well-defined, with galaxies more widely distributed over a 
span of $\sim 8$\,Mpc in the perpendicular direction. 
There is a denser, well-defined clump or filament of galaxies to
the southeast (grey points in Figure~\ref{fig_fil011}) that may be part of
a nearly north-south filament, but
at a perpendicular distance of $6.4$\,Mpc (if extrapolated to the north;
grey line in the figure) 
its association with the absorber is even less likely.  Taking the
dashed best-fit lines as the nearest filaments' ``centers,''
A1 and A2 lie at perpendicular distances of 4.7 and 2.7\,Mpc from the
centers of their respective galaxy structures.  

Although large-scale galaxy filaments are expected to trace substantial
reservoirs of WHIM gas, such X-ray absorbers are also likely to be
present in individual galaxy halos and low-mass groups 
\citep[e.g.][]{williams05}.  A1 is relatively isolated:
one galaxy lies a projected 588\,kpc to the southwest, but this is well 
outside the typical extent of warm-hot galaxy halos
\citep[up to 100\,kpc; e.g.][]{anderson10}.  
On the other hand, two moderately bright ($\sim 0.4$ and $0.8L_\star$) spiral
galaxies lie somewhat
closer to A2: 364 and 397 projected kpc to the east and 
southeast, respectively.  While it is still unlikely that gas associated
with these individual galaxies' halos extends to this distance, due to the 
proximity
of these galaxies to each other (343\,kpc projected separation and exactly
the same redshift) it is possible that they reside in a bound system 
comparable to the Local Group.
Such a system may contain diffuse bound warm-hot gas (analogous
to intracluster media) extending to several hundred kpc, which could
be responsible for the observed X-ray absorption.  However, the redshifts
of the galaxies ($z=0.0285\pm 0.001$) are at the upper edge of our selection
window and $+450$\,km\,s$^{-1}$ higher than the nominal absorber velocity; 
the median redshift of the galaxies in the larger filament is thus 
a marginally better match to that of the absorber.


\section{Discussion} \label{sec_discussion}
\subsection{Semantic uncertainties} \label{sec_caveats}
Due to their rather subjective definitions, the nature of these large-scale 
structures (e.g. filaments, sheets, and/or clumps), and which galaxies
should be included as members of the structures, are difficult to 
quantify \citep[for one novel method, see][]{bond10}.  Because they span 
$\sim 20$\,Mpc or more, simply counting galaxies
in a cylinder of this size around an absorber provides no information 
on the spatial configuration of the nearby large-scale structure, or 
whether a well-defined filament (if one is within the field) actually 
intercepts the absorber's position.  Similarly, traditional measures
of local galaxy density ($n$th-nearest neighbor, for example) are in a 
sense ``too local'' to provide information on WHIM-scale structures.

We therefore employ a semiqualitative technique of picking out
the nearest linear filamentary structures by eye, excluding galaxies which
lie significantly outside those structures, and defining the filament's
``center'' through a linear fit to the sky positions of the
galaxies in that redshift slice.  This combines the particular strengths
of human pattern recognition \citep[successfully employed by
other studies, e.g.~Galaxy Zoo;][]{lintott08} while still allowing quantitative
measures of the filaments' physical properties (dimensions, redshift
distribution, number of constituent galaxies, and distance from the
absorber to the center).  In principle these observations can then
be compared directly to simulations as long as comparable analysis
techniques are used for the mock galaxy catalogs.


\subsection{IGM, galaxy, or coincidence?}
\begin{figure}
\plotone{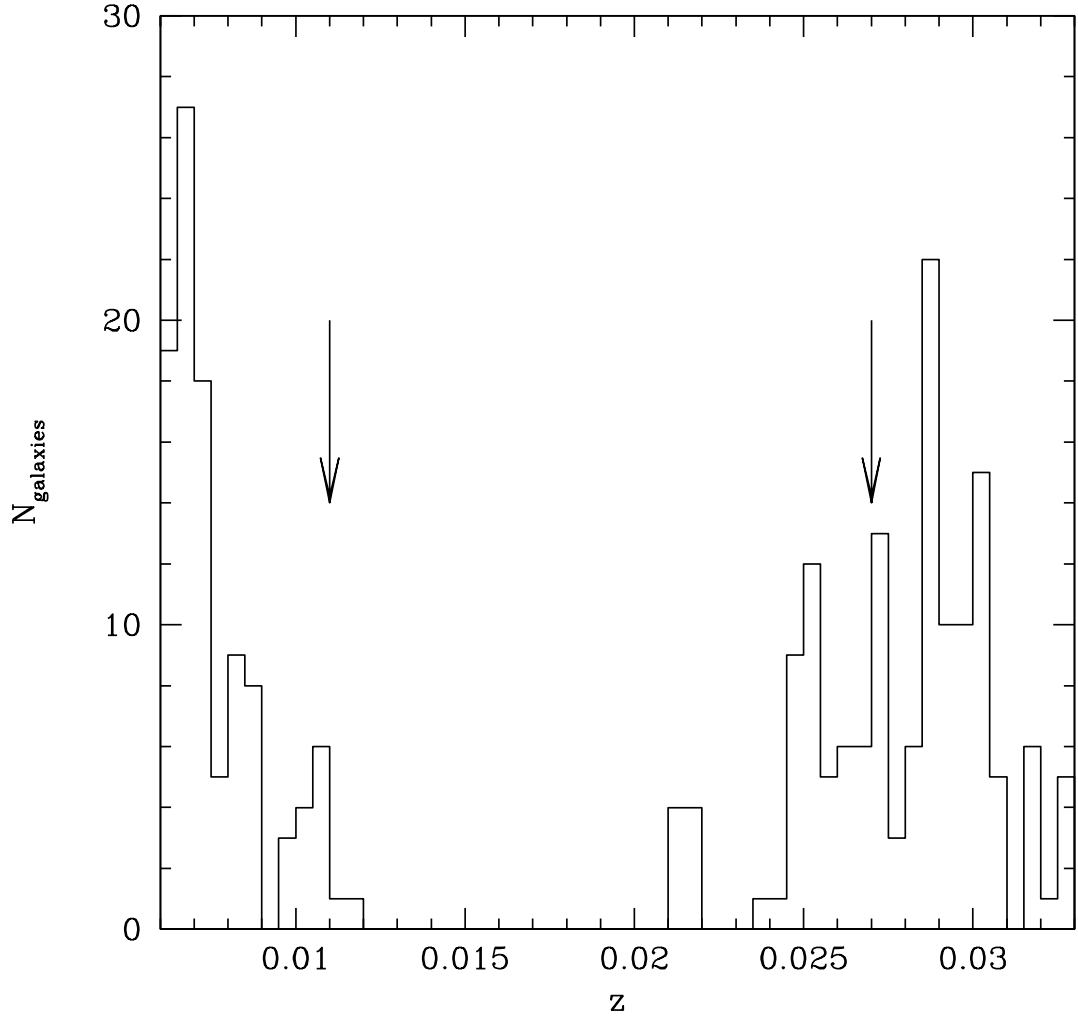}
\caption{Redshift histogram of SDSS galaxies within a $r=5$\,Mpc
cylinder centered on Mrk 421, spanning $z=0.005$ to $z=0.033$
(i.e. to just above the AGN redshift).  Arrows mark the two absorption
system redshifts reported by N05.
\label{fig_zhist}}
\end{figure}

\begin{figure}
\plotone{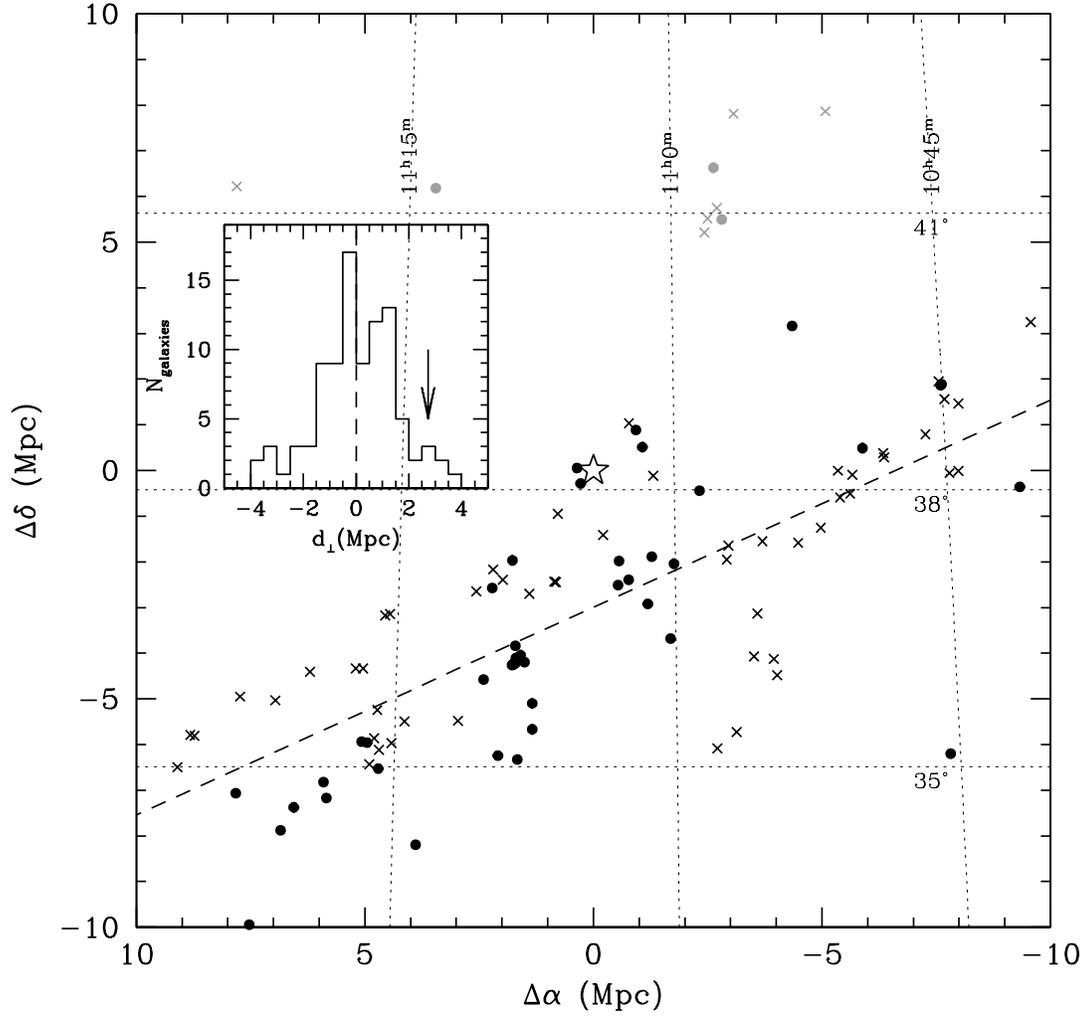}
\caption{Same as Figure~\ref{fig_fil011}, but for the $z=0.027$
absorber (A2).  Here the dashed line is a fit to galaxies with 
$\Delta\delta<5$\,Mpc, excluding the grey points to the north.  
\emph{Inset:} Distribution of $d_\perp$ 
from the center of the nearby filament, with the arrow 
showing the absorber's distance of 2.7\,Mpc.  In this case a
filamentary structure of at least $\sim 20$\,Mpc length is clearly
evident; the two galaxies near the center are $\sim 380$ projected
kpc from the absorber. 
\label{fig_fil027} }
\end{figure}

As noted previously, X-ray absorption systems like those reported by N05
can result from a variety of physical scenarios, including WHIM
filaments, individual galaxy halos or outflows, and intragroup gas.
Taken in its entirety, the evidence that the A2 absorption system is
associated with a large-scale galaxy filament is compelling.  The
galaxies in the vicinity of A2 fall almost exclusively
within a clear filamentary structure with length at least 20\,Mpc and
FWHM of 3.2\,Mpc (Figure~\ref{fig_fil027}).  The 95 galaxies in this 
structure have a median
redshift of $z=0.0269$, exactly coincident with the absorber, and are
more or less isotropically distributed around this redshift 
(with 52 galaxies at $0.0255<z<0.0270$ and 43 at $0.270<z<0.0285$).
This structure also defines a strong, narrow redshift spike 
(Figure~\ref{fig_zhist}), indicating that it is compact in
both width and the $\Delta z$ direction.  Although some of
the absorption could in principle arise in the intragroup medium of 
a galaxy pair $\sim 380$ projected kpc away, its redshift ($z=0.0285$) is not
as good a match to the absorption system, so its physical distance could
be as much as several Mpc.

On the other hand, the lower-redshift absorber A1 is not as clearly
associated with a large-scale structure.  While a possible filamentary
structure or sheet appears in Figure~\ref{fig_fil011}, its projected distance
is larger (4.7\,Mpc) and surface density lower than the filament near A2;
according to Figure~\ref{fig_zhist} fewer galaxies fall
near the absorber in redshift space as well.  Nor are
the galaxy redshifts in this structure in as good agreement with the 
absorber redshift
as A2 and its filament: 14 galaxies lie above $z=0.0110$ and 48 lie below,
with a median $z=0.0105$ (though this is still formally within the absorber's
redshift uncertainty).  No individual galaxy or group 
lies within 0.5\,Mpc of the absorber.  Given the relative weakness
of the A1 detection (seen in two $>2\sigma$ X-ray absorption
lines, as opposed to the four significant lines tracing A2), it is possible 
that this detection is spurious \citep[see also][]{kaastra06}, in which
case the lack of a nearby galaxy or large-scale structure would not
be surprising.  However, if the A1 absorber is
real, it may represent either a true ``void'' absorption system or 
probe gas associated with the rather distant, diffuse sheet of galaxies
to the north.

Given the ubiquity of filamentary and planar structures on $\sim 20$\,Mpc 
scales \citep{bond10}, the significance of these absorber-large scale
structure associations depend strongly on how common such structures
are around any given point -- for example, if a diffuse sheet
of galaxies like that toward A1 is typical in most $20\times 20$\,Mpc 
fields, the presence of such a sheet near A1 becomes less physically meaningful.
Conversely, if such sheets are rare, A1 and the sheet are almost certainly
associated.
To test this empirically, we drew 140 random samples
of galaxies from a larger portion of the SDSS ($\alpha=09^{\rm h}-16^{\rm h}$,
$\delta=10\degr-40\degr$; $z=0.026-0.030$) in boxes with dimensions identical
to Figures~\ref{fig_fil011} and \ref{fig_fil027}, smoothed the galaxy 
distributions with a $\sigma=1.25$\,Mpc Gaussian filter to better bring out 
extended structure, and searched by eye
for similar filamentary structures near the box centers.  

\begin{figure}
\plotone{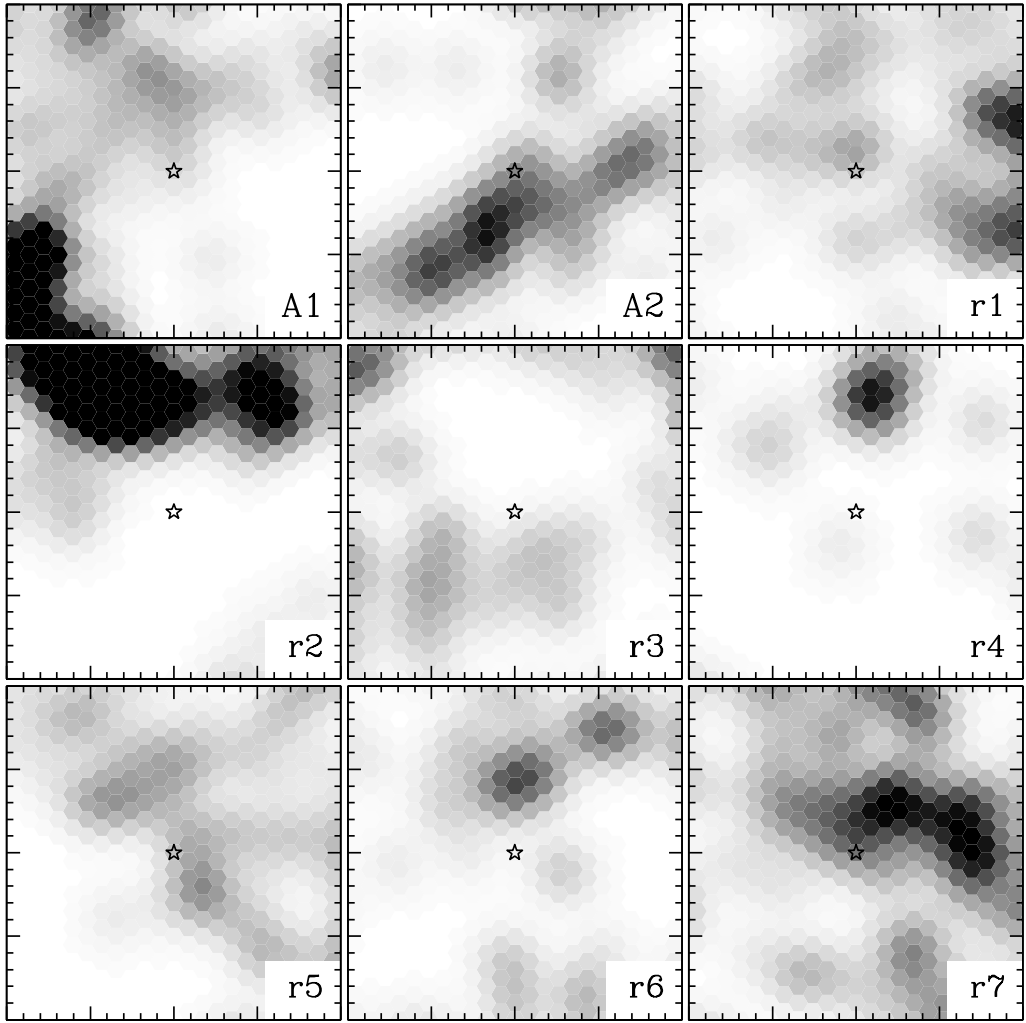}
\caption{Galaxy distributions around the two absorbers (A1, A2;
field sizes are identical to Figures~\ref{fig_fil011} and 
\ref{fig_fil027}) along with seven representative ``random slices'' from SDSS
($r1-r7$), smoothed with a $\sigma=1.25$\,Mpc Gaussian kernel.  Out of
140 such random realizations, $5-10$\% of these mock sightlines lie within 
filaments (panel $r7$) of comparable or higher density than that 
seen near A2, but a much higher fraction ($30-40$\%) fall in
less-concentrated regions like A1 (e.g.~$r1,r3,r5$).
\label{fig_random} }
\end{figure}

Seven such ``random'' samples are shown in Figure~\ref{fig_random} alongside 
smoothed versions of Figures~\ref{fig_fil011} and \ref{fig_fil027} (labeled 
as ``A1'' and ``A2'') for comparison.  Again the caveat
about the difficulty of quantifying such structures applies; however,
as many as $30-40$\% of the random samples exhibit relatively diffuse galaxy 
distributions like that near A1.
By contrast, strong filaments like the one coincident with A2 are rare:
only in $\sim 5-10$\% of these 140 random samples do structures comparable
to the filament in Figure~\ref{fig_fil027} 
($N_{\rm gal}\ga 100$, length $\sim 20$\,Mpc, width $\sim 5$\,Mpc)
fall within a few Mpc of the box center.  
This test further reinforces the conclusion that A2 is real and
associated with a nearby galaxy filament; on the other hand, the presence
of a relatively common galaxy sheet near A1 may be a coincidence.

\section{Summary} \label{sec_summary}
Using the SDSS spectroscopic survey, we have investigated the presence
and nature of large-scale galaxy structures in the vicinity of two 
possible WHIM  absorbers reported by
\citep{nicastro05a} in the \chandra\ spectrum of Mrk 421.  Our conclusions 
are as follows:
\begin{enumerate}
\item{The X-ray absorber at $z=0.027$ coincides with
a strong filamentary structure of galaxies.  Given that such structures
are expected to be tracers warm-hot intergalactic gas, this absorption
system may therefore trace the IGM associated with this filament.}
\item{A weaker filament or sheet of galaxies may be associated
with the weaker $z=0.011$ absorption system.  However, given the lower
significance of this X-ray detection, the larger distance to 
the galaxy structure (4.7\,Mpc), and the greater likelihood ($\sim 30-40$\%) 
of a chance superposition, this absorber-filament alignment may be
spurious.}
\item{Both of these absorbers lie more than 350\,kpc from the nearest
SDSS-detected galaxy; thus, if the X-ray detections are real, they likely
trace gas associated with large-scale structures rather than individual 
galaxies.}
\end{enumerate}

\acknowledgments
The authors thank the anonymous referee for constructive comments and
suggestions.  Partial support for this work was provided by 
NSF grant AST-0707417.

Funding for the SDSS and SDSS-II has been provided by the Alfred P. 
Sloan Foundation, the Participating Institutions, the National Science 
Foundation, the U.S. Department of Energy, the National Aeronautics and 
Space Administration, the Japanese Monbukagakusho, the Max Planck Society, 
and the Higher Education Funding Council for England. The SDSS Web Site is 
\url{http://www.sdss.org/}.

The SDSS is managed by the Astrophysical Research Consortium for the 
Participating Institutions. The Participating Institutions are the American 
Museum of Natural History, Astrophysical Institute Potsdam, University of 
Basel, University of Cambridge, Case Western Reserve University, University 
of Chicago, Drexel University, Fermilab, the Institute for Advanced Study, 
the Japan Participation Group, Johns Hopkins University, the Joint 
Institute for Nuclear Astrophysics, the Kavli Institute for Particle 
Astrophysics and Cosmology, the Korean Scientist Group, the Chinese 
Academy of Sciences (LAMOST), Los Alamos National Laboratory, the
Max-Planck-Institute for Astronomy (MPIA), the Max-Planck-Institute for 
Astrophysics (MPA), New Mexico State University, Ohio State University, 
University of Pittsburgh, University of Portsmouth, Princeton University,
the United States Naval Observatory, and the University of Washington.

\begin{thebibliography}

\bibitem[Abazajian et al.(2009)]{abazajian09} Abazajian, K., et al.~2009, 
\apjs, 182, 543
\bibitem[Anderson \& Bregman(2010)]{anderson10} Anderson, M.~E.~\& Bregman, 
J.~N.~2010, \apj, 714, 320
\bibitem[Blanton et al.(2001)]{blanton03} Blanton, M.~R., et al.~2003, 
\apj, 592, 819
\bibitem[Bond et al.(2010)]{bond10} Bond, N.~A., Strauss, M.~A., \&
Cen, R.~2010, \mnras, 406, 1609
\bibitem[Buote et al.(2009)]{buote09} Buote, D., et al.~2009, \apj, 695, 1351
\bibitem[Cen \& Ostriker(1999)]{cen99} Cen, R.~\& Ostriker, J.~P.~1999,
\apj, 514, 1
\bibitem[Dav\'e et al.(2001)]{dave01} Dav\'e, R., et al.~2001, \apj, 552, 473
\bibitem[Dav\'e et al.(2010)]{dave10} Dav\'e, R., et al.~2010, \mnras, 
in press (arXiv:1005.2421)
\bibitem[Fang et al.(2002)]{fang02} Fang, T., Marshall, H.~L., Lee, J.~C.,
Davis, D.~S., \& Canizares, C.~R.~2002, \apj, 572, L127
\bibitem[Fang et al.(2007)]{fang07} Fang, T., Canizares, C.~R., \& Yao, 
Y.~2007, \apj, 670, 992
\bibitem[Fang et al.(2010)]{fang10} Fang, T., et al.~2010, \apj, 714, 1715
\bibitem[Galeazzi et al.(2009)]{galeazzi09} Galeazzi, M., Gupta, A., \&
Ursino, E.~2009, \apj, 695, 1127
\bibitem[Kaastra et al.(2006)]{kaastra06} Kaastra, J.~S., et al.~2006, \apj,
652, 189
\bibitem[Lintott et al.(2008)]{lintott08} Lintott, C.~J., et al.~2008,
\mnras, 389, 1179
\bibitem[Mannucci et al.(2007)]{mannucci07} Mannucci, F., Bonnoli, G.,
Zappacosta, L., Maiolino, R., \& Pedani, M.~2007, \aap, 468, 807
\bibitem[Nicastro et al.(2005a)]{nicastro05a} Nicastro, F., et al.~2005, \apj, 629, 700 (N05)
\bibitem[Nicastro et al.(2005b)]{nicastro05b} Nicastro, F., et al.~2005, 
Nature, 433, 495
\bibitem[Rasmussen et al.(2007)]{rasmussen07} Rasmussen, A., et al.~2007,
\apj, 656, 129
\bibitem[Strauss et al.(2002)]{strauss02} Strauss, M.~A., et al.~2002, 
\aj, 124, 1810
\bibitem[Werner et al.(2008)]{werner08} Werner, N., et al.~2008, \aap,
482, L29
\bibitem[Williams et al.(2005)]{williams05} Williams, R.~J., et al.~2005,
\apj, 631, 856
\bibitem[Williams et al.(2006)]{williams06} Williams, R.~J., Mathur, S.,
Nicastro, F., \& Elvis, M.~2006, \apj, 642, L95
\bibitem[Williams et al.(2007)]{williams07} Williams, R.~J., Mathur, S.,
Nicastro, F., \& Elvis, M.~2007, \apj, 665, 247
\bibitem[York et al.(2000)]{york00} York, D.~G., et al.~2000, \aj, 120, 1579
\bibitem[Zappacosta et al.(2005)]{zappacosta05} Zappacosta, L., Maiolino, R.,
Mannucci, F., Gilli, R., \& Schuecker, P.~2005, \mnras, 357, 929
\bibitem[Zappacosta et al.(2010)]{zappacosta10} Zappacosta, L., et al.~2010,
\apj, 717, 74
\end{thebibliography}
\end{document}